\documentclass{elsart}

 \usepackage{epsfig}

\usepackage{amssymb}

\begin{document}

\begin{frontmatter}

\markboth{Tiwari, Rahaman \& Ray}{Five Dimensional Cosmological
Models in General Relativity}

\title{Five Dimensional Cosmological Models in General Relativity}

\author{R. K. Tiwari}\footnote{Department of Mathematics, Government Model Science College, Rewa 486 001, M.P., India \\E-mail:
rishitiwari59@rediffmail.com}.
\author{Farook Rahaman}\footnote{Department of Mathematics, Jadavpur University, Kolkata 700 032, W.B., India \\E-mail:
farook\_rahaman@yahoo.com}.
\author{Saibal Ray}\footnote{Department of Physics, Government College of Engineering \& Ceramic Technology, Kolkata 700 010,
W.B., India \\E-mail: saibal@iucaa.ernet.in}

\maketitle

\begin{abstract}
A Five dimensional Kaluza-Klein space-time is considered in the
presence of a perfect  fluid source with variable G and $\Lambda$.
An expanding universe is found by using a relation between the
metric potential and an equation of state. The gravitational
constant is found to decrease with time as $G \sim
t^{-(1-\omega)}$ whereas the variation for the cosmological
constant follows as $\Lambda \sim t^{-2}$, $\Lambda \sim (\dot
R/R)^2$ and $\Lambda \sim \ddot R/R$ where $\omega$ is the
equation of state parameter and $R$ is the scale factor.
\end{abstract}

\end{frontmatter}

                                                                                                                            ¨

\section{Introduction}
In recent years there has been considerable interest in the
cosmological models with variable gravitational constant $G$ and
the cosmological constant $\Lambda$. Variation of the
gravitational constant was first suggested by Dirac
\cite{Dirac1937} in an attempt to understand the appearance of
certain very large numbers, when atomic and cosmic worlds are
compared. He postulates that the gravitational constant $G$
decreases inversely with cosmic time. Canuto et al.
\cite{Canuto1977a,Canuto1977b} made numerous suggestions based on
different arguments that $G$ is indeed time dependent. Beesham
\cite{Beesham1986} has studied the creation with variable $G$ and
pointed out the variation of the form $G \sim t^{-1}$, originally
proposed by Dirac \cite{Dirac1937}. On the other hand, Einstein
introduced the cosmological constant $\Lambda$ to account for a
stable static universe as appeared to him at the time. When he
later knew of the universal expansion he regretted its inclusion
in his field equations. Now cosmologist believe that is not
identically, but very close to zero. They relate this constant to
the vacuum energy that first inflated a universe causing it to
expand \cite{Zeldovich1968}. From the point of view of particle
physics a vacuum energy could correspond to quantum field that is
diluted to its present small value. However, other cosmologists
dictate a time variation of this constant in order to account for
its present smallness \cite{Overduin1998}. The variation of its
constant could resolve some of the standard model problems like
$G$, the constant $\Lambda$ is a gravity coupling and both should
therefore be treated on an equal footing.

The generalized Einstein's theory of gravitation with time
dependent $G$ and $\Lambda$ has been proposed by Lau
\cite{Lau1985}. The possibility of variable $G$ and $\Lambda$ in
Einstein theory has also been studied by Dersarkissian
\cite{Dersarkissian1985}. This relation plays an important role in
cosmology. Berman \cite{Berman1991a} and Sistero
\cite{Sistero1991} have considered the Einstein field equations
with perfect fluid and variable $G$ and for Robertson-Walker line
element. Kalligas et al. \cite{Kalligas1992} have studied FRW
models with variable $\Lambda$ and $G$ and discussed the possible
connection with power-law time dependence of $G$. Abdussattar and
Vishwakarama \cite{Abdussattar1997} presented R-W models with
variable $\Lambda$ and $G$ by admitting a contracted Ricci
collineation along the fluid flow vector. Recently some of us and
others have studied cosmological models with variable $G$ and
$\Lambda$ in a diversified fields
\cite{Pradhan2002,Khadekar2006,Ray2007a,Ray2007b,Pradhan2007,Khadekar2007,Singh2007,Tiwari2008,Singh2008a,Singh2008b,Tiwari2009a,Tiwari2009b}.
Thus the implication of time varying $\Lambda$ and $G$ are
important to study the early evolution of the universe.

In the present paper a five dimensional Kaluza-Klein cosmological
model is considered with variable $G$ and $\Lambda$ which provides
an expanding universe. It is found that the gravitational constant
decreases with time as $G \sim t^{-(1-\omega)}$ whereas the
cosmological constant decreases as $\Lambda \sim t^{-2}$, $\Lambda
\sim (\dot R/R)^2$ and $\Lambda \sim \ddot R/R$ where $\omega$ is
the equation of state parameter, $R$ is the scale factor. However,
we have come across with some awkward situations in connection to
the deceleration parameter and the age of the Universe as far as
the scenario of present accelerating Universe is concerned. In the
present investigation, our approach is similar to that of Collins
et al. \cite{Collins1977} for solving the Einstein field
equations. The paper is organized as follows: In Sections 2 and 3
we present the basic field equations governing the models and
their solutions. A critical discussion and conclusion are provided
in Section 4.

\section{The Einstein field equations for the cosmological model}
We consider five dimensional Kaluze-Klein space-time given
by
\begin{eqnarray}
ds^{2} = dt^{2} - A^{2}(dx^2 + dy^2 + dz^2) - B^2 d\psi^{2}
\end{eqnarray}
where $A$ and $B$ are function of the temporal coordinate $t$
only.

The universe is assumed to be filled with distribution of matter
represented by energy-momentum tensor for a perfect fluid
\begin{eqnarray}
 T_{ij}= (\rho +p)v_{i} v_{i} - pg_{ij}
\end{eqnarray}
where $\rho$ is the energy density of the cosmic matter and $p$ is
its pressure, $v_{i}$  is the unit flow vector such that
$v_{i}v^{i} =1$.

We assume that the matter content obeys an equation of state
\begin{eqnarray}
p=\omega\rho, ~0 \leq \omega \leq 1.
\end{eqnarray}

The field equations are those of Einstein but with time dependent
cosmological and gravitational constants and are given  by
Weinberg \cite{Weinberg1972}
\begin{eqnarray}
R^{ij} - \frac{1}{2}Rg^{ij} = 8\pi GT^{ij} - \Lambda g^{ij}.
\end{eqnarray}
The spatial average scale factor $R(t)$ is given by
\begin{eqnarray}
R^4=A^3B
\end{eqnarray}
and average volume scale factor $V=R^4$.

The  average Hubble parameter H may be generalized  in anisotropic
cosmological model as
\begin{eqnarray}
H=\frac{1}{4} \dot X=\dot Y=\frac{1}{4}\left(\frac{3\dot A}{A} +
\frac{\dot B}{B}\right)
\end{eqnarray}
where $X=logV$ and $Y=log R$ and an over head dot denotes ordinary
differentiation with respect to cosmic time $t$. We also have
\begin{eqnarray}
H=\frac{1}{4} (H_1+H_2+H_3+H_4)
\end{eqnarray}
where $H_1=H_2=H_3=\dot A/A$ and $H_4=\dot B/B$ are Hubble's
factor in the directions of $x$, $y$, $z$ and $\psi$ respectively.

For the line element (1) the Einstein field equations (4) yield
the following equations
\begin{eqnarray}
\frac{3\dot A^2}{A^2} + \frac{3\dot A \dot B}{AB}=8\pi G \rho +
\Lambda,
\end{eqnarray}
\begin{eqnarray}
\frac{2\ddot A}{A} + \frac{\ddot B}{B}+ \frac{2\dot A \dot
B}{AB}+\frac{\dot A^2}{A^2}=-8\pi G p + \Lambda,
\end{eqnarray}
\begin{eqnarray}
\frac{3\ddot A}{A}+\frac{3\dot A^2}{A^2}=-8\pi G p + \Lambda.
\end{eqnarray}
If we use the equivalent energy conservation of general relativity
by taking the covariant derivative of Einstein field equations (4)
we find that
\begin{eqnarray}
\dot \rho+(\rho+p)\left(\frac{3\dot A}{A} + \frac{\dot
B}{B}\right) + \frac{\dot G}{G} \rho + \frac{\dot \Lambda}{8 \pi
G} = 0.
\end{eqnarray}
This provides \cite{Pradhan2007,Singh2008b}
\begin{eqnarray}
\dot \rho+(\rho+p)\left(\frac{3\dot A}{A} + \frac{\dot
B}{B}\right) =0
\end{eqnarray}
as well as
\begin{eqnarray}
\dot \Lambda = - 8 \pi \dot G \rho
\end{eqnarray}
which will be satisfied by the solutions of the field equations as
can be verified easily.

\section{The general solutions to the field equations}
The equations (3), (8) - (10) and (12) are five independent
equations with six unknowns $A$, $B$, $\rho$, $p$, $G$ and
$\Lambda$. Hence to get a realistic solution we assume that the
expansion scalar $\theta$ in the model is proportional to the
shear $\sigma$ \cite{Singh2008a}. This condition leads to the
relation between metric potential \cite{Collins1977}
\begin{eqnarray}
B = A^n
\end{eqnarray}
where $n (\neq 1)$ is a constant.

From equations (9) and (10), we have
\begin{eqnarray}
\frac{\dot A}{A} - \frac{\dot B}{B}=\frac{k_1}{A^3B}
\end{eqnarray}
where $k_1$ is a constant of integration. Using equation (14) in
equation (15) and then integrating it we get the lime element (1)
as
\begin{eqnarray}
ds^{2} = dt^{2} - \left[k_1 \frac{(n+3)}{(1-n)}t +
k_2(n+3)\right]^{2/(n+3)}(dx^2 + dy^2 + dz^2) \nonumber \\
-\left[k_1 \frac{(n+3)}{(1-n)}t + k_2(n+3)\right]^{2n/(n+3)}
d\psi^{2}.
\end{eqnarray}
By suitable changes of constants the line element (1) can be
written as
\begin{eqnarray}
ds^{2} = dt^{2} - (at + b)^{2/(n+3)}(dx^2 + dy^2 + dz^2) -
(at + b)^{2n/(n+3)} d\psi^{2}
\end{eqnarray}
where $a=k_1(n+3)/(1-n)$ and $b=k_2(n+3)$ are two positive constants.

For the model (17), the spatial volume $V$, matter density $\rho$,
pressure $p$, gravitational parameter $G$, cosmological parameter
$\Lambda$ are given by
\begin{eqnarray}
V= R^4 = (at + b),
\end{eqnarray}
\begin{eqnarray}
\rho=\frac{k_2}{(at + b)^{(1+\omega)}},
\end{eqnarray}
\begin{eqnarray}
p=\frac{\omega k_2}{(at + b)^{(1+\omega)}},
\end{eqnarray}
\begin{eqnarray}
G=\frac{3a^2(n+1)}{4\pi k_2(1+\omega)(n+3)^2 (at + b)^{(1-\omega)}},
\end{eqnarray}
\begin{eqnarray}
\Lambda=\frac{3(1-\omega)a^2(n+1)}{(1+\omega)(n+3)^2(at + b)^2}.
\end{eqnarray}

The expansion scalar $\theta$ and shear $\sigma$ are
\begin{eqnarray}
\theta=3H=\frac{a}{(at + b)},
\end{eqnarray}
\begin{eqnarray}
\sigma=\frac{\alpha}{(at + b)}
\end{eqnarray}
where $\alpha$ is a positive constant.

The critical density $\rho_c$, vacuum density $\rho_{\Lambda}$ and density parameter
$\Omega$ are given by
\begin{eqnarray}
\rho_c=\frac{3H^2}{8\pi G}=\frac{(1+\omega)(n+3)^2}{32k_2(n+1)(at
+ b)^{(1+\omega)}},
\end{eqnarray}
\begin{eqnarray}
\rho_{\Lambda}=\frac{\Lambda}{8\pi G}=\frac{k_2(\omega-1)}{2(at +
b)^{(1+\omega)}},
\end{eqnarray}
\begin{eqnarray}
\Omega=\frac{8\pi G\rho}{3H^2}=\frac{(n+1)}{(1+\omega)(n+3)^2}.
\end{eqnarray}

Keeping in mind the form of the solution (18), i.e. $R^4 =(at +
b)$, one can express all the above parameters as function of the
scale factor $R$ as well which we will follow in the following two
sub-cases.

\subsection{Special case studies}

\subsubsection{The radiation era}

For the radiation epoch, we have $\omega=1/3$. In this case
\begin{eqnarray}
\rho=k_2 R^{-4/3},
\end{eqnarray}
\begin{eqnarray}
G=\frac{9a^2(n+1)}{16\pi K_2(n+3)^2}R^{-2/3},
\end{eqnarray}
\begin{eqnarray}
\Lambda=\frac{-3a^2(n+1)}{2(n+3)^2}R,
\end{eqnarray}
\begin{eqnarray}
\rho_{\nu}=\frac{-k_2}{3} R^{-4/3},
\end{eqnarray}
\begin{eqnarray}
\rho_c=\frac{(n+3)^2}{24(n+1)} R^{-4/3}.
\end{eqnarray}

We observe that in this phase matter density is three times the vacuum density.
The cosmic density parameter $\Omega$ in this phase is
\begin{eqnarray}
\Omega =\frac{3(n+1)}{4(n+3)^2}.
\end{eqnarray}

\subsubsection{The dust epoch}

For the phase dominated by dust matters, we have $\omega =0$. In this epoch
\begin{eqnarray}
\rho=k_2 R^{-1},
\end{eqnarray}
\begin{eqnarray}
G=\frac{3a^2(n+1)}{4\pi k_2(n+3)^2}R^{-1},
\end{eqnarray}
\begin{eqnarray}
|\Lambda|=\frac{3a^2(n+1)}{(n+3)^2}R^{-2},
\end{eqnarray}
\begin{eqnarray}
\rho_{\nu}=\frac{k_2}{2} R^{-1},
\end{eqnarray}
\begin{eqnarray}
\rho_c=\frac{(n+3)^2}{32(n+1)} R^{-1}.
\end{eqnarray}
\begin{eqnarray}
\frac{|\Lambda|}{8\pi G\rho} =\frac{1}{2}.
\end{eqnarray}

\section{Discussion and conclusion}
In the present phenomenological study five dimensional
Kaluza-Klein cosmological models with varying $G$ and $\Lambda$
are obtained. The models obtained present an expansion scalar
$\theta$ bearing a constant ratio to the anisotropy in the
direction of space like unit vector $\lambda^{\prime}$ and with
equation of state $p=\omega\rho, ~0 \leq \omega \leq 1$. If we
inspect the solution set then it can be observed from the equation
(18) that the spatial volume $V$ is zero at $t = t_o$ where $t_o =
- b/a$ and expansion scalar $\theta$ is infinite at $t =t_o$ which
shows that the Universe starts evolving with zero volume and
infinite rate of expansion at $t = t_o$. The scale factors also
vanish at $t = t_o$ and hence the model has a {\it point type}
singularity at the initial epoch. Since $\sigma/\theta$  is
constant, as evident from the solutions (23) and (24), the
anisotropy does not die out asymptotically. The model admits a
negative $\Lambda$ unless $\omega=1$ and these type of models for
the Universe with negative $\Lambda$ are available in the
literature \cite{Vishwakarma2005}. We observe from the general
solution set that for the stiff fluid ($\omega=1$) the model
reduces to standard Kaluza-Klein type Zel'dovich universe
\cite{Zeldovich1972} with $\Lambda = 0$, $G =$  constant and $\rho
=p \sim t^{-2}$ for the limiting case $b \rightarrow 0$ (Figs. 1
\& 2). At $t= t_o$ limit all the parameters $G$, $\Lambda$,
$\sigma^2$ and $\theta$ are infinite in value, whereas all these
become zero as $t \rightarrow \infty$. The ratio between cosmic
matter density parameter $\Omega_m (= 8\pi G\rho/3H^2)$ and cosmic
vacuum-energy density parameter $\Omega_{\Lambda} (=\Lambda/3H^2)$
scales as
\begin{eqnarray}
\frac{\Omega_m}{\Omega_{\Lambda}} = \frac{\Lambda}{8\pi G\rho}=\frac{\omega-1}{2},
\end{eqnarray}
whereas cosmic vacuum-energy density parameter can be given as
\begin{eqnarray}
\Omega_{\Lambda}=\frac{\Lambda}{3H^2}=\frac{q(n+1)(\omega-1)}{(n+3)^2(1+\omega)}.
\end{eqnarray}

The above two expressions related to cosmic matter and  vacuum density parameters
do vanish for the stiff fluid condition $\omega=1$.

If we look at the solution expressed in the equation (21) then we
can observe that in general $G \sim t^{-(1-\omega)}$. Thus for the
dust case ($\omega =0$) this case is similar to the result of
Dirac \cite{Dirac1937}, Canuto et al.
\cite{Canuto1977a,Canuto1977b} and Beesham \cite{Beesham1986}. On
the other hand, equations (22) and (23) at once imply that
$\Lambda \sim H^2$ for any constant values of $\omega$ and $n$.
Again, from the solutions (18) and (22) one can find $\Lambda \sim
\ddot R/R$. On the other hand for the limiting case $b \rightarrow
0$ the equation (22) yields the relation $\Lambda \sim t^{-2}$.
Thus we obtain $\Lambda \sim H^2 = (\dot R/R)^2$, $\Lambda \sim
\ddot R/R$ and $\Lambda \sim t^{-2}$ which is in accordance with
the main dynamical laws one finds in literature proposed for the
decay of $\Lambda$. Note that we find all these phenomenological
relations for $\Lambda$ directly from the solution set which
generally consider as ad hoc basis in cosmological research. The
dynamical law $\Lambda \sim (\dot R/R)^2$, has been proposed by
Carvalho et al. \cite{Carvalho1992} and considered by Salim and
Waga \cite{Salim1993}, Arbab and Abdel-Rahman \cite{Arbab1994a},
Wetterich \cite{Wetterich1995}, Arbab \cite{Arbab1994b} and very
recently by Ray and his collaborators as well as Pradhan and his
collaborators
\cite{Ray2007a,Ray2007b,Ray2007c,Khadekar2007,Ray2009,Mukhopadhyay2010a}.
The decay law of the form $\Lambda \sim \ddot R/R$ has been
considered by Arbab \cite{Arbab2003}, Khadekar et al.
\cite{Khadekar2006}, Ray et al. \cite{Ray2007a}, Ray and
Mukhopadhyay \cite{Ray2007c} and Ray et al. \cite{Ray2009}. The
dynamical law $\Lambda \sim t^{-2}$ has been considered by several
authors, e.g. Bertolami \cite{Bertolami1986}, Berman and Som
\cite{Berman1990a}, Berman \cite{Berman1991b}, Beesham
\cite{Beesham1994}, Pradhan et al. \cite{Pradhan2002} to mention a
few. We observe that $|\Lambda|$ decays faster than $G$ where as
$p$, $\rho$, $\rho_{\nu}$, $\rho_c$ scales $R^{-(\omega +1)}$ as
can be seen from the Figs. 1 - 4. In the model we see that the
quantity $G$ satisfies the condition for a Machian cosmological
solution i.e. $G\rho \sim H^2$ \cite{Berman1990b}.

However, as far as the scenario of present accelerating Universe
is concern our models seems suffer a lot in connection to the
different physical parameters, such as the deceleration parameter
and the age of the Universe. The deceleration parameter $q$ for
the present model is $q_0 = 3$ and hence does not correspond to
the observational result $q_0 \approx -0.5$
\cite{Tripp1997,Sahni1999}. However, we have observed that for the
scale factor of the form $a(t) = t^n$ we can have a negative
deceleration parameter when $n>1$. In our present case $n = 1/4$
which yields $q = +ve$. Here the age of the universe is given by
$t_0=\frac{1}{3} H_0^{-1}$ which is also differs from the present
estimate $t_0= H_0^{-1} \approx 14$ Gyr
\cite{Hansen2002,Sievers2003,Tegmark2003,Spergel2006}. This value,
therefore, obviously suffers from the low age problem
\cite{Mukhopadhyay2010b} as even the oldest globular clusters show
a value which is $t \approx 11$ Gyr
\cite{Chaboyer1998,Krauss2003}. In this connection it is argued
that dark energy models without $\Lambda$ suffer from low age
problem \cite{Sahni2000,Mukhopadhyay2010b} though even in the
presence of $\Lambda$ we are not in a hopeful position. One of the
reasons of these discrepancies may be due to the constant nature
of the equation of state parameter $\omega$ which can not
accommodate with the present accelerating phase of the Universe.
However, this needs further investigation.

\section*{Acknowledgments}
The authors (FR \& SR) is thankful to the authority of
Inter-University Centre for Astronomy and Astrophysics, Pune,
India for providing Visiting Associateship under which a part of
this work was carried out. We all are thankful to the referee for
valuable suggestions which have enabled us to improve the
manuscript substantially.

{}

\begin{figure*}
\begin{center}
\vspace{0.5cm} \psfig{file=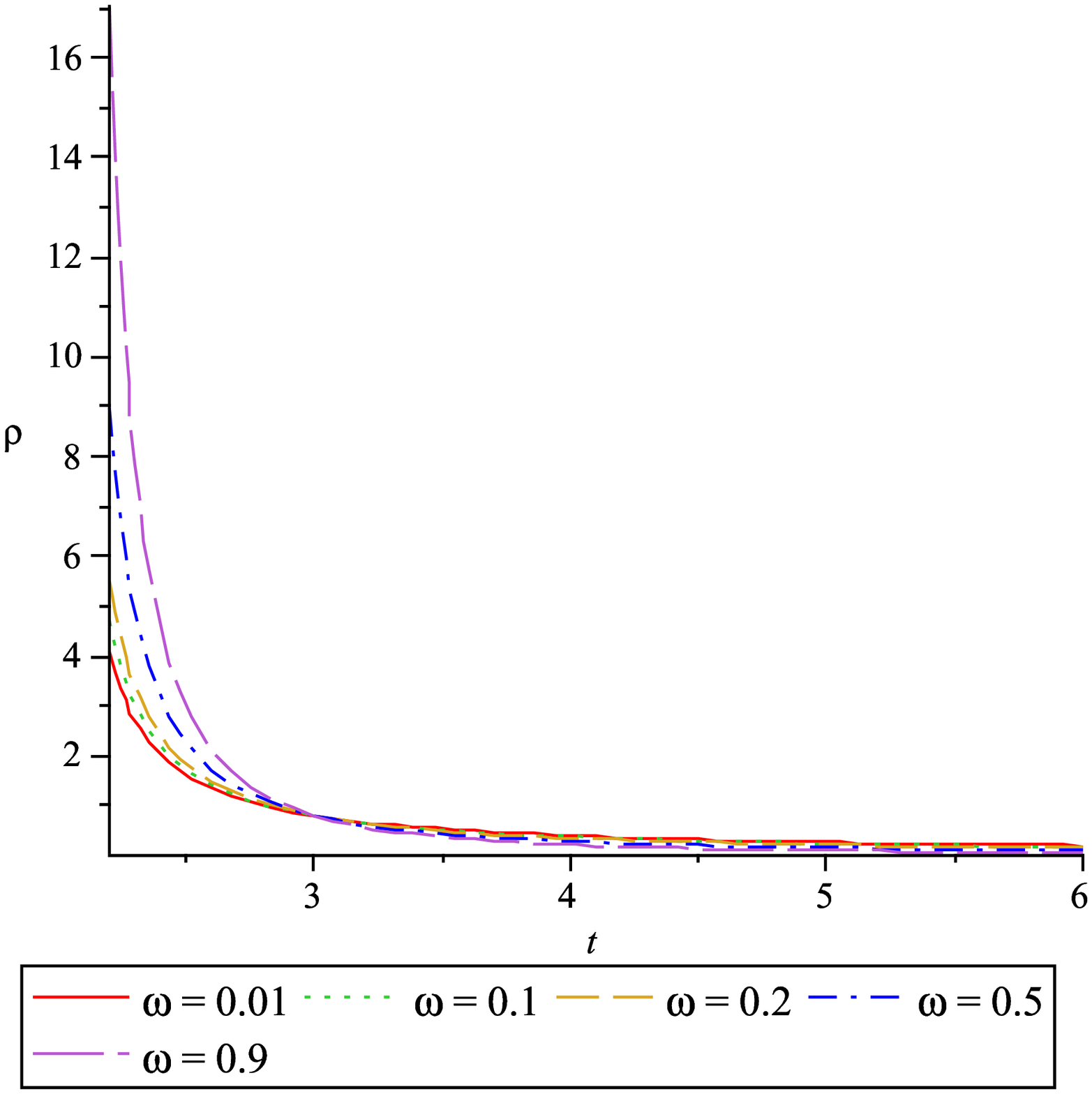,width=0.6\textwidth}
\caption{Variation of matter density with time for the given
values of $\omega$ under the constraint $k_2 = 0.8$, $a = 1$ and
$b=-2$.}
 \label{fig1}
\end{center}
\end{figure*}

\begin{figure*}
\begin{center}
\vspace{0.5cm} \psfig{file=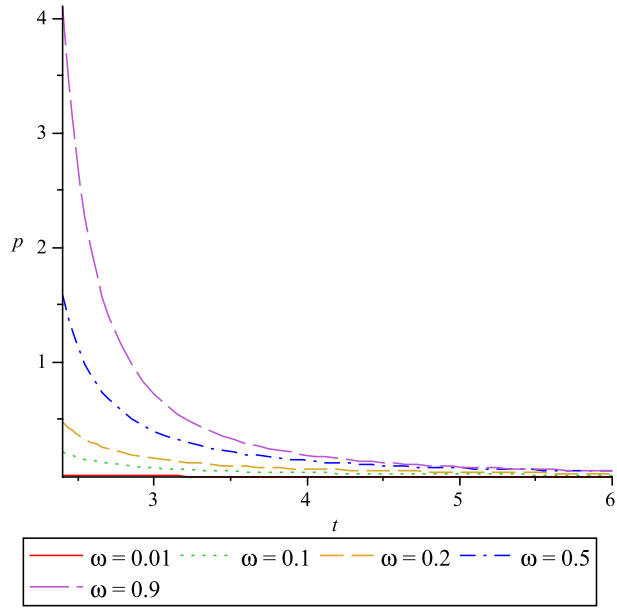,width=0.6\textwidth}
\caption{Variation of pressure with time for the given values of $\omega$.}
 \label{fig2}
\end{center}
\end{figure*}

\begin{figure*}
\begin{center}
\vspace{0.5cm} \psfig{file=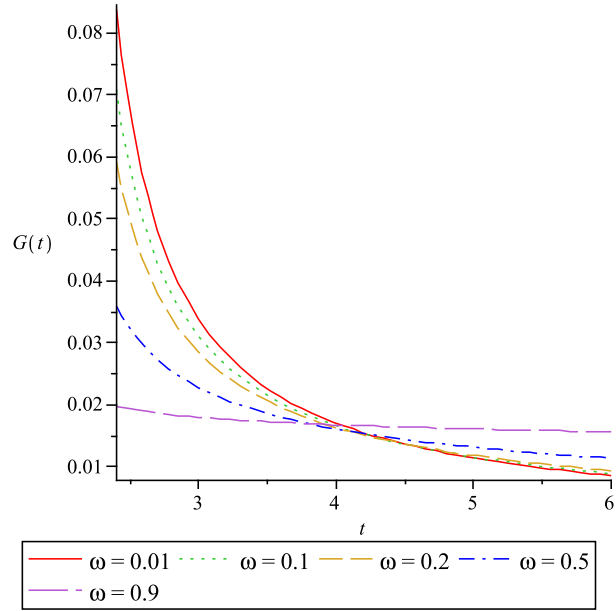,width=0.6\textwidth}
\caption{Variation of time-dependent gravitational constant with time for the given values of $\omega$.}
 \label{fig3}
\end{center}
\end{figure*}

\begin{figure*}
\begin{center}
\vspace{0.5cm} \psfig{file=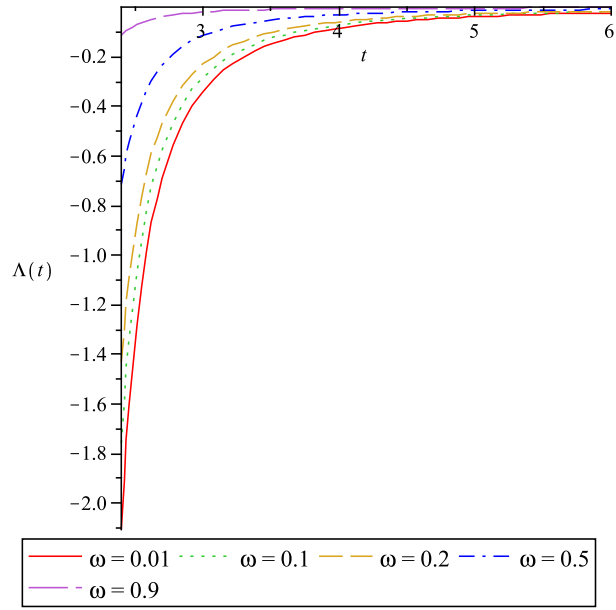,width=0.6\textwidth}
\caption{Variation of time-dependent cosmological constant with time for the given values of $\omega$.}
 \label{fig4}
\end{center}
\end{figure*}

\end{document}